\documentclass[aps]{revtex4}%
\usepackage{amsfonts}
\usepackage{amsmath}
\usepackage{amssymb}
\usepackage{graphicx}%
\begin{document}
\title{Is Negative Refraction, Perfect Focusing Compatible with Quantum Mechanics?}
\author{P. R. Berman}
\affiliation{Michigan Center for Theoretical Physics, FOCUS Center, and Physics Department,
University of MIchigan, Ann Arbor, Michigan 48109-1120}
\date{\today}

\begin{abstract}
In light of experiments in atom optics, the compatibilty of negative
refraction, perfect focusing with quantum mechanics is brought into question.

\end{abstract}
\maketitle

There has been renewed interest in media having negative index of refraction
\cite{neg}. The idea that a medium having $\epsilon<0$, $\mu<0$ can act as a
perfect lens for a point particle was proposed by Veselago \cite{neg} and
Pendry \cite{pend}, and analyzed recently by Merlin \cite{merl}. Experimental
evidence for the focusing effect has been reported recently \cite{exp}.
Ziolkowski and Heyman \cite{ziol} carried out analytical and numerical
calculations to see if perfect focusing could be achieved or approximated.
They showed that, in principle, perfect focusing could be achieved for a
loss-less, dispersionless medium if $\epsilon=\mu=-1$. In all their
simulations, ideal focusing was not achieved and they claim that it would be
impossible to do so for any realistic material. I would like to question the
idea of perfect focusing of a point source from a somewhat different
perspective. Imagine one has an atom interferometer and excites a
\textquotedblright single\textquotedblright\ atom that is in two arms of the
interferometer. The excited atom can then undergo spontaneous decay. Such an
experiment was carried out by Chapman et al. \cite{pritch}. They found that if
excitation occurred when the spatial wave function of the atom in the two arms
of the interferometer was separated by less than an optical wavelength, the
interference pattern was not destroyed. On the other hand, if the separation
was greater than an optical wavelength, the atom interference pattern was
washed out. These experimental results are explained in terms of
\ \textquotedblright which path\textquotedblright\ information, assuming that
it is impossible to localize an atom by the radiation it emits to better than
an optical wavelength. Now imagine the light emitted from such atoms was
incident on a perfect lens (or a pair of such lenses placed above and below
the plane of the interferometer), such as that proposed by Veselago and
Pendry. Since the lens serves as a perfect lens for point sources,
\textquotedblright which path\textquotedblright\ information could be obtained
even for atoms separated by less than a wavelength. In this case, the atom
interference pattern in the interferometer would be destroyed even for path
separations much less than an optical wavelength. Such an effect is not
observed experimentally. Note that it is not necessary to have the lens
actually present in the experiment. The interference pattern is destroyed if,
\emph{in principle}, such a measurement can be made. Other processes involving
cooperative emission from atoms confined to less than an optical wavelength
also rely on this path indistinguishability and that path indistinguishability
would no longer be guaranteed if \textquotedblright perfect\textquotedblright%
\ lenses were available. Thus the existence of a perfect lens for point
particles appears to be inconsistent with experimental results in atom optics
and quantum mechanics.

This work is supported by the U. S. Office of Army Research under Grant No.
DAAD19-00-1-0412, the National Science Foundation under Grant No. PHY-0244841,
the FOCUS Center Grant, and by the National Science Foundation through a grant
for the Institute for Theoretical Atomic and Molecular Physics at Harvard
University and Smithsonian Astrophysical Observatory. I would like to thank
the members of ITAMP for their hospitality during my visit. I am pleased to
acknowledge helpful discussion with M. Moore, R.\ Merlin, and G. Shvets.

\end{document}